\documentclass[a4paper,fleqn,usenatbib]{mnras}

\usepackage{amsmath}	% Advanced maths commands
\usepackage[T1]{fontenc}
\usepackage{ae,aecompl}

\usepackage{graphicx}	% Including figure files

\usepackage{amssymb}	% Extra maths symbols

\title[Modern yields per stellar generation]{Modern yields per stellar generation: the effect of the IMF}
\author[F. Vincenzo et al.]{F. Vincenzo$^{1,2}$\thanks{E-mail:
vincenzo@oats.inaf.it}, F. Matteucci$^{1,2,3}$, F. Belfiore$^{4,5}$, R. Maiolino$^{4,5}$
\\
$^{1}$Dipartimento di Fisica, Sezione di Astronomia, Universit\`a di Trieste, via G.B. Tiepolo 11, 34100, Trieste, Italy\\
$^{2}$INAF, Osservatorio Astronomico di Trieste, via G.B. Tiepolo 11, 34100, Trieste, Italy\\
$^{3}$INFN, Sezione di Trieste, Via Valerio 2, 34100, Trieste, Italy\\
$^{4}$Cavendish Laboratory, University of Cambridge, 19 J. J. Thomson Avenue, Cambridge CB3 0HE, UK\\
$^{5}$Kavli Institute for Cosmology, University of Cambridge, Madingley Road, Cambridge CB3 0HA, UK}

\begin{document}

 \date{Accepted 2015 November 03. Received 2015 November 03; in original form 2015 March 26}

\pagerange{\pageref{firstpage}--\pageref{lastpage}} \pubyear{2015}

\maketitle

\label{firstpage}

%%%%%%%%%%%%%%%%%%%%%%%%%%%%%%%%%%%%%%
%%%%%%%%%%%%%%%%%%%%%%%%%%%%%%%%%%%%%%

\begin{abstract}
Gaseous and stellar metallicities in galaxies are nowadays routinely
used to constrain the evolutionary processes in galaxies. This requires the
knowledge of the average yield per stellar generation, $y_{\text{Z}}$, i.e. 
the quantity of metals that a stellar population releases into the interstellar
medium (ISM), which is generally assumed to be a fixed fiducial value.
Deviations of the observed metallicity from the expected value of $y_{\text{Z}}$ are
used to quantify the effect of outflows or inflows 
of gas, or even as evidence for biased metallicity calibrations or
inaccurate metallicity diagnostics. 
Here we show that $\rm y_{\text{Z}}$ depends significantly on the Initial Mass Function
(IMF), varying by up to a factor larger than three, for the range of IMFs typically
adopted in various studies.
Varying the upper mass cutoff of the IMF implies a further variation of $y_{\text{Z}}$ by an additional factor 
that can be larger than two. 
These effects, along with the variation of the gas mass fraction restored into the ISM by
supernovae ($R$, which also depends on the IMF), may yield to deceiving results, if not 
properly taken into account. In particular, metallicities
that are often considered unusually high can actually be explained in terms of yield 
associated with commonly adopted IMFs such as the Kroupa (2001) or Chabrier (2003). 
We provide our results for two different sets of stellar yields (both affected by 
specific limitations) finding that the uncertainty introduced by this assumption can be as large as $\sim0.2$ dex. 
Finally, we show that $y_{\text{Z}}$ is not substantially affected by the initial stellar metallicity as long as 
$\text{Z}> 10^{-3}~\text{Z}_{\odot}$. 
\end{abstract}


\section{Introduction}

\label{introduction} 

The analysis of the chemical enrichment of galaxies is a powerful tool to constrain galaxy evolutionary 
processes. The content of metals in galaxies, both in the interstellar medium (ISM) and in stars, 
depends critically on the past star formation history and on the net effect of outflows and inflows, which are 
some of the key mechanisms in shaping galaxy evolution. In order to extract valuable information from the 
observed metallicities, it is crucial to compare them with the amount of metals expected 
to be produced by the integrated star formation. To achieve this is necessary to have accurate 
information on the amount of each chemical element injected into the ISM by each type of star, 
i.e. the so-called stellar yields. Generally, most observations 
provide information only on the global content of metals, or on a single chemical element which is 
taken as representative of the global metallicity (by assuming that, for instance, the 
abundance of the various elements scales proportionally to the solar relative abundances). 
Moreover, many studies do not deal with the relative delayed enrichment of different chemical species. 
Therefore, the quantity that is often used is the so-called average 
{\it yield per stellar generation}, or {\it net yield} (generally indicated as $y_{\text{Z}}$, or simply $y$), 
which is the total mass of metals that a stellar population releases into the ISM, 
normalized to the mass locked up into low-mass (long-lived) stars and stellar remnants.

%%%%%%%%%%%%%%%%%%%%%%%%%%%%%%%%%%%%%%%%%%

     \begin{figure}
\includegraphics[width=9.3cm]{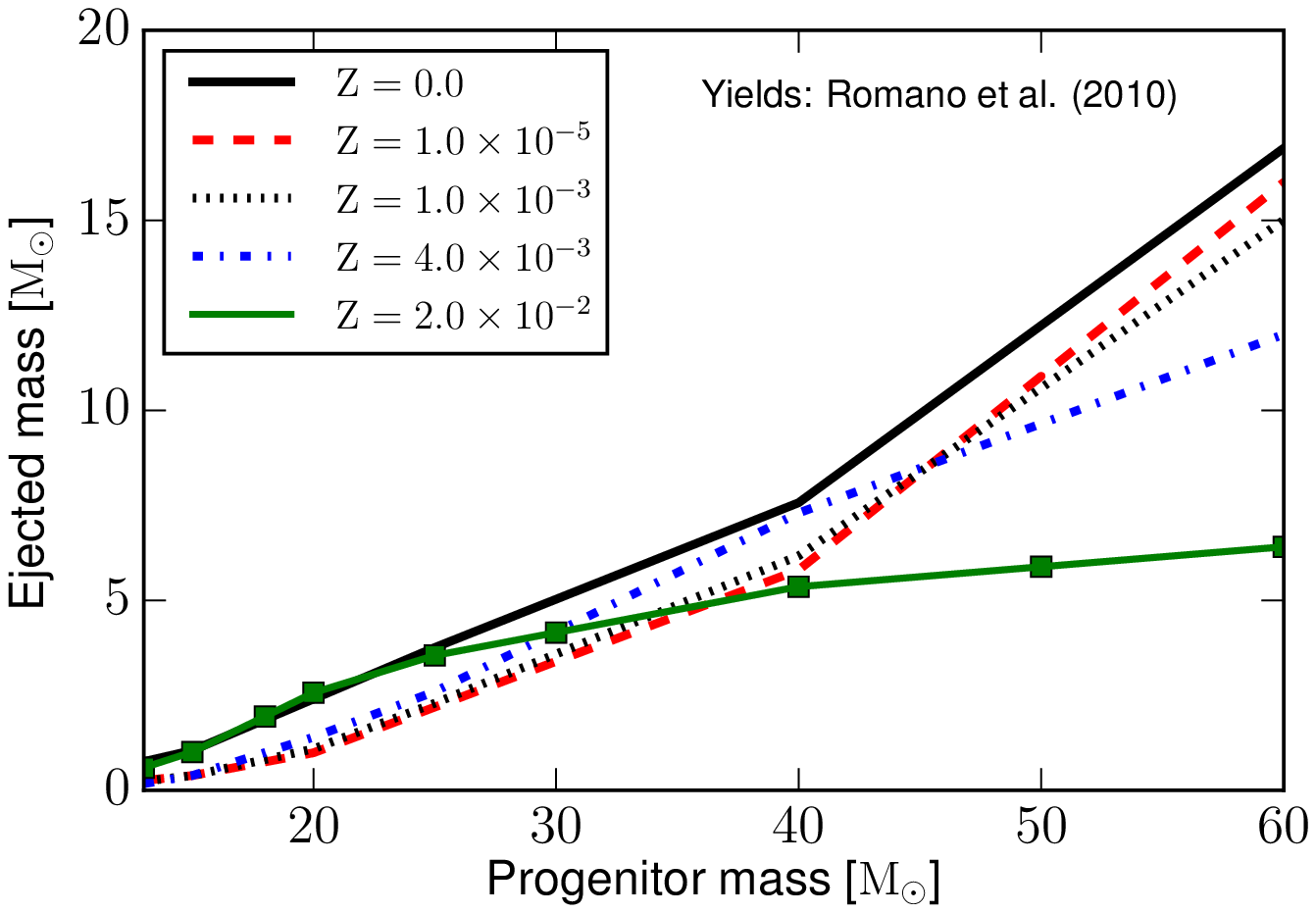}
\caption{In this figure, we report the ejected mass of oxygen as a function of the progenitor mass, for different initial stellar metallicities. This set of stellar yields is 
from \citet{romano2010}. The solid line in black corresponds to the stellar yields at $\text{Z}=1.0\times10^{-10}$; 
the dashed line in red to $\text{Z}=1.0\times10^{-5}$; 
the black dotted line to $\text{Z}=1.0\times10^{-3}$; the dashed-dotted line in blue to $\text{Z}=3\times10^{-3}$, and the green squares to the stellar yields at 
$\text{Z}=2.0\times10^{-2}$. }
     \label{fig:Oyields}
   \end{figure} 
   
%%%%%%%%%%%%%%%%%%%%%%%%%%%%%%%%%%%%%%%%%%

\par Historically, this approach was first used in the early work of \citet{searle1972}, who derived the
relation between gas phase metallicity Z and gas fraction $\mu=M_{\text{gas}} / (
M_{\text{gas}}+M_{\star} ) $ for a closed box model (see \citealt{tinsley1980}, or
\citealt{matteucci2001} for a detailed analysis): \begin{equation} \label{eq:simple} \text{Z}=y_{\text{Z}}\ln \left( \frac{1}{\mu} \right). \end{equation} 
This simple model is based upon the assumptions that the galaxy is one-zone and closed; the initial gas
mass is of primordial chemical composition; the initial mass function (IMF) is invariant, and the mixing
of the gas in the galaxy is always instantaneous and complete, and that metals are instantaneously
recycled for the formation of the new generation of stars. The latter is dubbed ``instantaneous recycling
approximation'' (IRA, \citealt{tinsley1980}). Within this simplified (but widely used) approach the
further approximation is that  all stars with $m\geq1\,\text{M}_{\sun}$ die instantaneously, while all stars with $m<1\,\text{M}_{\sun}$ 
have infinite lifetime; this way, the effect of stellar lifetimes in the equations can be neglected and
the effect subsumed into a net \textit{return fraction} ($R$). 

\par Although the IRA assumption is strong, 
it still represents a good approximation for those chemical elements produced and restored into the ISM 
by stars with short lifetimes. The best example of such a chemical element is oxygen, which is also representative 
of the global metallicity Z, since it is the most abundant heavy element by mass. On the other hand, the ISM evolution of chemical elements 
produced by long-lifetime sources cannot be followed by analytical models working with the IRA assumption. 
Examples of such chemical elements are nitrogen and carbon, which are mainly synthesized by low- and intermediate-mass stars (LIMS), and iron, 
mainly produced by Type Ia SNe. To take into account the stellar lifetimes with a high level of detail, numerical models 
of chemical evolution should be used (see \citealt{matteucci2012}). 

%%%%%%%%%%%%%%%%%%%%%%%%%%%%%%%%%%%%%%%%%%

     \begin{figure}
\includegraphics[width=9.3cm]{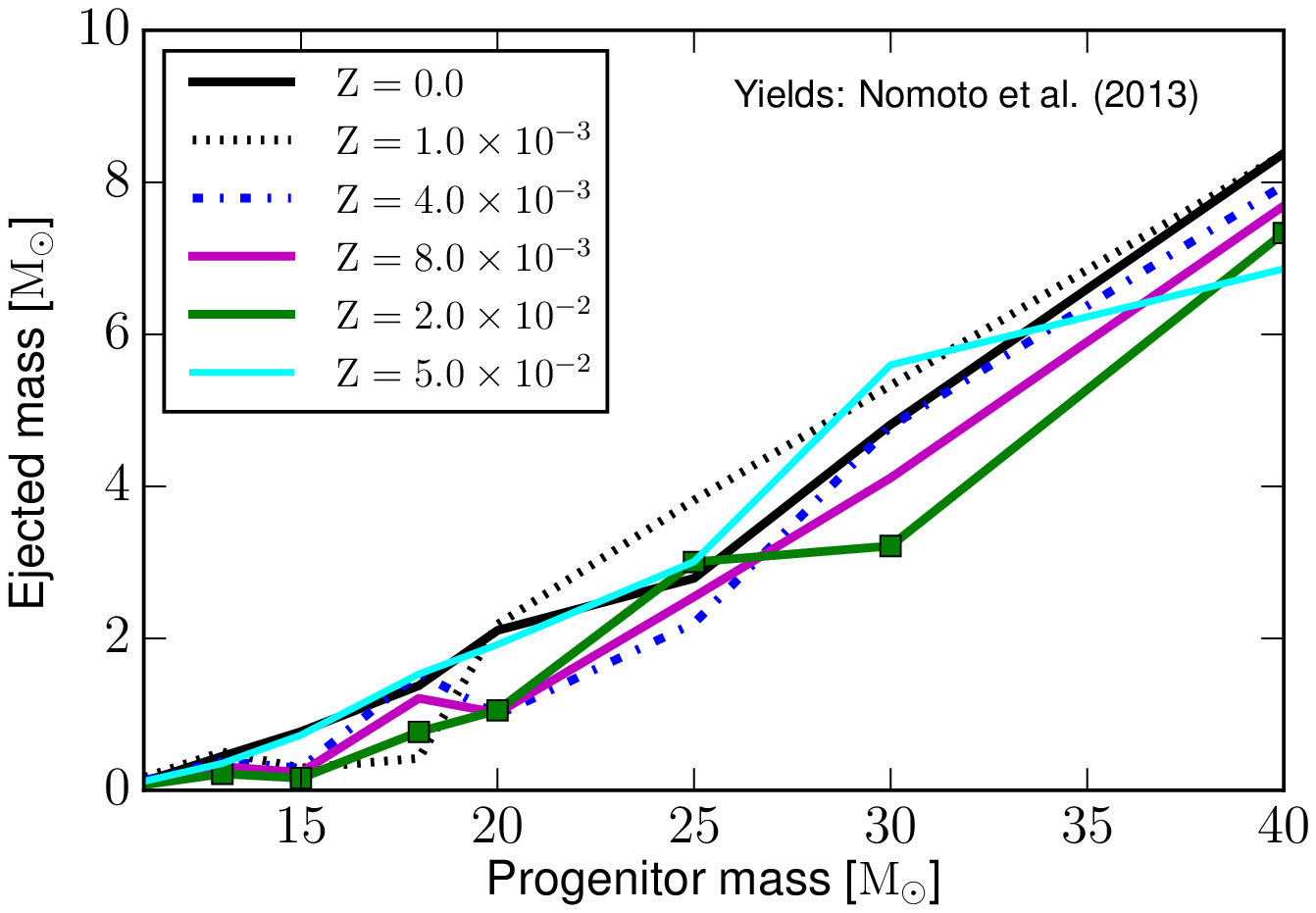}
\caption{ In this figure, we show how ejected mass of oxygen vary as a function of the progenitor mass, for different initial stellar metallicities, when assuming the 
\citet{nomoto2013} stellar yields. The solid line in black corresponds to the stellar yields at $\text{Z}=0.0$; the dotted line in black to $\text{Z}=1.0\times10^{-3}$; 
the dashed-dotted line in blue to $\text{Z}=4.0\times10^{-3}$; 
the solid line in magenta to $\text{Z}=8.0\times10^{-3}$; the solid line in green to $\text{Z}=2.0\times10^{-2}$, and the solid line in cyan to the stellar yields at 
$\text{Z}=5.0\times10^{-2}$. }
     \label{fig:Oyields_nomoto}
   \end{figure} 
   
%%%%%%%%%%%%%%%%%%%%%%%%%%%%%%%%%%%%%%%%%%

\par The yield per stellar generation is a key in the context of analytical models of chemical evolution, even in the most complex ones, 
which include the effect of outflows and inflows, as well as variations of the
star formation efficiency, i.e. normalization and slope of the relation between
gas mass and star formation rate, (see, for example, 
\citealt{bouche2010,spitoni2010,dekel2013,peng2014,zahid2014,ascasibar2014,spitoni2015}), as well as in numerical simulations
(e.g. \citealt{cole2000,delucia2004}). The comparison of these models with the extensive observations that
are providing metallicity measurements for large samples of galaxies locally and at high redshift
(e.g. \citealt{savaglio2005,maiolino2008,troncoso2014,steidel2014})
enable us to provide important constraints on these various processes, {\it modulo an accurate knowledge of the yield per
stellar generation}.

\par In most of the studies, the yield per stellar generation is taken as a fixed value (typically
about $0.012$-$0.045$), with this value changing from work to work. 
For example, a net yield of oxygen $y_\mathrm{O}=0.015$ is assumed in \citet{peeples2014}, whereas  
$y_\mathrm{O}=5.7\times10^{-3}$ in \citet{zahid2014} and $y_\mathrm{O}=3.13\times10^{-3}$ in \citet{ho2015}; 
particularly high are the values $y_{\text{O}}=0.040$-$0.045$ assumed in \citet{delucia2004}, $y_{\text{O}}=0.03$ in \citet{croton2006}, and 
$y_{\text{O}}=0.04$ in \citet{bower2008}. 
However, since the net yield is a combination of yields from different stellar masses, 
it is clear that it must depend on the IMF. This has sometimes been acknowledged (e.g. \citealt{henry2000,kobayashi2011,ho2015}), 
but never really taken into consideration when using the net yield in the various models. 
In particular, several works derive the stellar mass and SFR by assuming a given IMF and then adopt a fiducial net yield that is derived from 
a completely different IMF. 
Moreover, there is
some evidence that the IMF may vary in different classes of galaxies. 
This implies that different yields
per stellar generation should be used. Finally, since the stellar nucleosynthetic yield have a metallicity dependence, 
it is important to check the effect of metallicity on the IMF-integrated net yield.

\par To tackle the issues presented above, in this article we calculate yields per stellar generation for the most commonly
adopted IMFs and investigate their metallicity dependence (although the latter effect is shown to be minor),
by comparing the results for two modern compilations of nucleosynthetic yields  \citep{romano2010,nomoto2013}, 
which have been throughly tested in the past against the best available data for galaxies, 
although we remark on the fact that each of them is still affected by specific limitations (as discussed in the following Sections).  
We mostly focus on the yield of oxygen, which is the element which is most commonly used as a tracer
of the global metallicity, and for which the IRA approximation is appropriate.
However, we will also provide the yield per stellar generation for the total mass of metals, although
this should be used with caution, given the enrichment delay of various elements (e.g. iron, nitrogen,
carbon, etc...), for which the IRA approximation is arguable. 

\par In Section \ref{definitions} we define the quantities we have computed in this work 
and specify the set of stellar yields which we have assumed and the IMFs which we have explored.
In Section \ref{results} we report and discuss our results. 
Finally, in Section \ref{conclusions} we summarize the main conclusions.

\section{Definitions and assumptions} \label{definitions}
 
 \subsection{Yield per stellar generation and return mass fraction}
 
We define as \textit{yield per stellar generation} 
$y_{i}(\text{Z})$, or \textit{net yield}, the ratio of the global gas mass in the form of a given chemical element $i$ newly produced and restored 
into the ISM by a simple stellar population with initial metallicity Z to the amount of mass locked up in low mass stars and stellar remnants 
\citep{tinsley1980,maeder1992,matteucci2001}: 
\begin{equation} 
\label{eq:y} y_{i}(\text{Z}) =
\frac{1}{1-R(\text{Z})}\frac{\int^{m_{\text{up}}}_{m_{\text{long-liv}}}m\,p_{i}(m,\text{Z})\,\phi(m)\,dm}{\int^{m_{\text{up}}}_{0.1~\text{M}_{\sun}}m\,\phi(m)\,dm}, 
\end{equation}
where: \begin{enumerate}       
       \item  $p_{i}(m,\text{Z})=\frac{M_{\text{ej},i}(m,\text{Z})}{m}$ is the so-called \textit{stellar yield}, which is defined such that $m\cdot{p_{i}(m,\text{Z})}$ 
       represents the mass in the form of the $i$-th chemical element  
       newly formed and ejected into the ISM by stars with initial mass $m$ and metallicity Z; 
       \item  $\phi(m)$ is the IMF, namely the mass-spectrum over which the stars of each single stellar generation are distributed at their birth; 
       \item $m_{\text{long-liv}}=1.0\,\text{M}_{\sun}$ is the maximum mass of the so-called long-lived stars, which do not
	   	pollute the ISM; 
       \item $m_{\text{up}}$ is the upper mass cutoff of the IMF; 
	   in our standard case, we assume $m_{\text{up}}=100\,\text{M}_{\sun}$, however in the second part of the article we will also investigate the effect of varying
	   $m_{\text{up}}$.  
       
      \end{enumerate}
Finally, $R$ represents the so-called \textit{return mass fraction}, which is defined as the total mass fraction (including both processed and unprocessed material) 
returned into the ISM by a stellar generation: \begin{equation}
             \label{eq:R} R(\text{Z}) =
			 \frac{\int^{m_{\text{up}}}_{m_{\text{long-liv}}}(m-M_{R}(m,\text{Z}))\,\phi(m)\,dm}{\int^{m_{\text{up}}}_{0.1~\text{M}_{\sun}}m\,\phi(m)\,dm},
            \end{equation}
with $M_{R}(m,\text{Z})$ being the mass of the stellar remnant left by a stars with initial mass $m$ and metallicity Z. 

%%%%%%%%%%%%%%%%%%%%%%%%%%%%%%%%%%%%%%%%%%

 \begin{table*}
 \caption[]{ \footnotesize{In this Table, we report the numerical values of $\langle P_{\mathrm{O}} \rangle$, which is defined as the IMF-averaged stellar yield of oxygen 
in the mass range $M=10$-$40\,\mathrm{M}_{\sun}$, for different metallicities Z. 
 'MM02' and 'HMM05' stand for \citet{meynet2002} and \citet{hirschi2005}, respectively, 
 which correspond to the results of the Geneva stellar models also included in \citet{romano2010}. 'NKT13' stands for \citet{nomoto2013}.
}}
 \begin{tabular}{c | c c c|c c c|c c c|c c c}
 \hline
 \small{Stellar yields} & \small{$v_\mathrm{rot}$ [km s$^{-1}$]} & Z & \small{$\langle P_{\mathrm{O}} \rangle$ } & \small{$\langle P_{\mathrm{O}} \rangle$ } 
 & \small{$\langle P_{\mathrm{O}} \rangle$ } & \small{$\langle P_{\mathrm{O}} \rangle$ } \\
 \hline
 \rule{0pt}{3ex}
  & & &  \citet{salpeter1955} & \citet{chabrier2003} & \citet{kroupa1993}  & \citet{kroupa2001} \\
\rule{0pt}{4ex}
  MM02 & $0$ & $4.0\times10^{-3}$       & $0.004$ & $0.007$ & $0.002$ & $0.007$   \\
 \rule{0pt}{2.5ex}   
    				            & $300$ &  & $0.006$ & $0.010$ & $0.003$ & $0.010$   \\
 \rule{0pt}{4ex}   
  HMM05    &  $0$ & $2.0\times10^{-2}$  & $0.006$ & $0.010$ & $0.003$ & $0.010$   \\
 \rule{0pt}{2.5ex}   
				             & $300$ &   & $0.009$ & $0.015$ & $0.005$ & $0.014$   \\
 \rule{0pt}{4ex}   
 NKT13   & no & $4.0\times10^{-3}$  & $0.007$ & $0.011$ & $0.004$ & $0.010$   \\
  \rule{0pt}{2.5ex}   
                & no & $2.0\times10^{-2}$ &  $0.006$ & $0.009$ & $0.003$ & $0.009$   \\
 \hline
\end{tabular}
\label{table_yields}
\end{table*} 

 %%%%%%%%%%%%%%%%%%%%%%%%%%%%%%%%%%%%%%%%%%

\par If one changes the quantity $m_{\text{long-liv}}$ in accordance to the age of the galaxy, 
then one would obtain equations very similar to the ones numerically solved by current models of chemical evolution 
(see, for example, \citealt{matteucci2012}). In principle, the assumption of  $m_{\text{long-liv}}=1.0\,\text{M}_{\sun}$ provides correct results only for 
stellar populations which are older than $\approx7.1\,\text{Gyr}$, which corresponds to the lifetime of an $1\,\text{M}_{\sun}$ star,
according to \citet{padovani1993}, although the lifetimes of low-mass stars can also be influenced by metallicity, particularly at very low $\text{Z}$ 
\citep{gibson1997}. 

   %%%%%%%%%%%%%%%%%%%%%%%%%%%%%%%%%%%%%%%%%%

  \begin{figure}
\includegraphics[width=9.3cm]{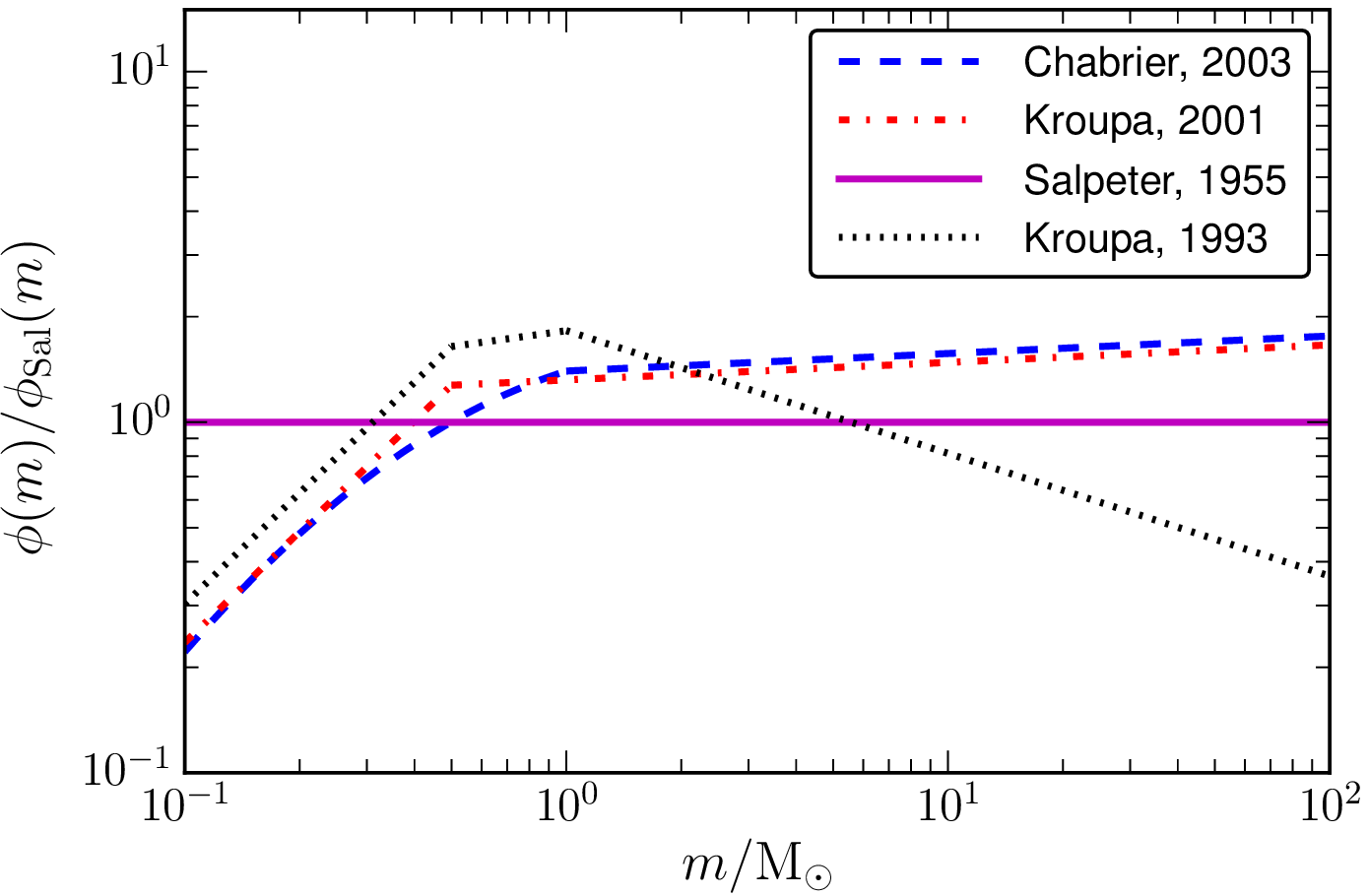}
\caption{In this figure, we show the trend of the different IMFs studied in this work, as normalized with respect to the \citet{salpeter1955} IMF. 
The dotted line in black corresponds to the \citet{kroupa1993} IMF; the dotted-dashed line in red to the \citet{kroupa2001} IMF; the dashed line in blue 
to the \citet{chabrier2003} IMF, and the solid line in purple corresponds to the \citet{salpeter1955} IMF. 
}
     \label{fig:IMF}
   \end{figure} 
   
%%%%%%%%%%%%%%%%%%%%%%%%%%%%%%%%%%%%%%%%%%

  %%%%%%%%%%%%%%%%%%%%%%%%%%%%%%%%%%%%%%%%%%

  \begin{figure}
\includegraphics[width=9.3cm]{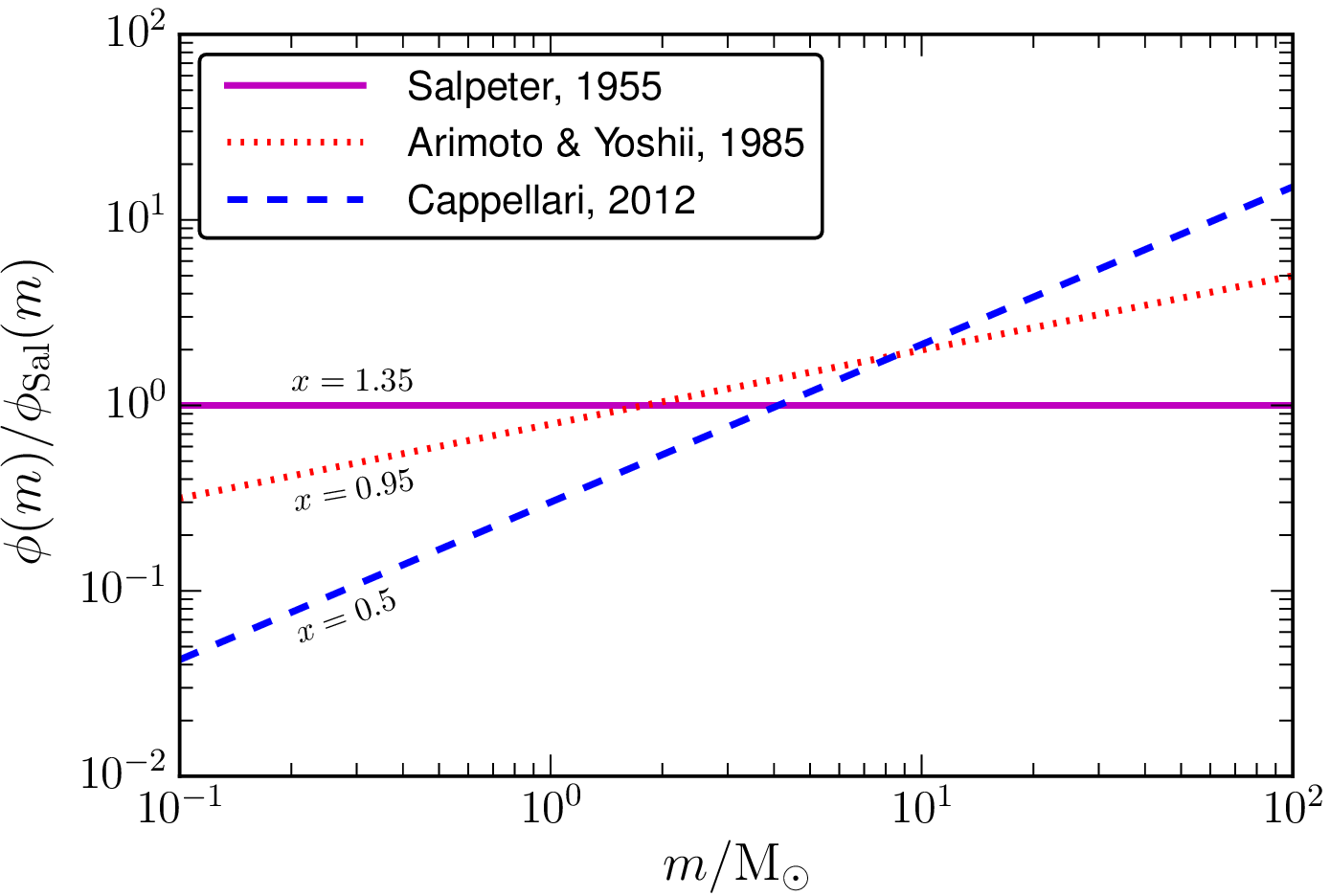}
\caption{In this figure, we compare the \citet[blue dashed line]{cappellari2012} top-heavy IMF with the IMFs of \citet[red dotted line]{arimoto1987} and 
\citet[solid line in magenta]{salpeter1955}. As in Fig. \ref{fig:IMF}, all the IMFs are normalized with respect to the \citet{salpeter1955} IMF. }
     \label{fig:IMF_topheavy}
   \end{figure} 
 
 %%%%%%%%%%%%%%%%%%%%%%%%%%%%%%%%%%%%%%%%%%

 \subsection{Stellar yields and initial mass function}
 
 \par We use a numerical code of chemical evolution to explore the effect of the metallicity and IMF on the yields of oxygen per stellar generation and 
 on the return mass fraction.   
We provide our results for the following sets of stellar yields: 
\begin{enumerate}
\item the set provided by \citet{romano2010}, 
which assume the stellar yields of \citet{karakas2010} for LIM stars, and the He, C, N and O stellar yields 
of the Geneva stellar models for massive stars \citep{meynet2002, hirschi2005,hirschi2007,ekstrom2008}; for heavier elements, 
which are not relevant for this study, \citet{romano2010} assume the stellar yields of massive stars of \citet{kobayashi2006}; 
\item the stellar yields of \citet{nomoto2013}, which include the stellar yields of LIM stars of \citet{karakas2010}, and the stellar yields of \citet{nomoto2006}, 
\citet{kobayashi2006}, \citet{kobayashi2011}, Tominaga et al. (unpublished) for core-collapse supernovae (SNe). 
\end{enumerate} The mass of the stellar remnants have been collected by \citet{romano2010} from the work of \citet{kobayashi2006}. 
Nevertheless, according to the \citet{talbot1973} formalism, 
we compute the return mass fraction with the \citet{romano2010} stellar yields, 
by summing the ejecta of all the chemical elements (both the processed and the unprocessed ones) for each stellar mass, and this quantity turns out to be dominated by 
the H and He contributions.

\par The \citet{romano2010} compilation of stellar yields for He, C, N and O include the results of models which take into account the combined effect of mass loss 
and rotation (see also \citealt{maeder2009} for a detailed discussion), whereas the \citet{nomoto2013} stellar yields have been computed by models which do 
include standard mass loss but not the effect of rotation. With standard mass loss, only C and N have been lost before supernova explosions. However, the mass 
loss driven by rotation 
turns out to be particularly important at almost solar metallicity and above in depressing the oxygen stellar yields of the 
most massive stars ($M\ga30$-$40\,\text{M}_{\sun}$, see also Fig. \ref{fig:Oyields}). In fact, mass loss increases with stellar metallicity and stars of high metal content 
loose H, He, but also C, through radiatively line driven winds. Therefore, the C production is increased by mass loss whereas the oxygen production is decreased, 
since part of C which would have been transformed into O, is lost from the star \citep{maeder1992}. Finally, the effect of rotation is to produce mixing and 
enhances mass loss, with the efficiency of the mixing process being larger at lower metallicities  (see also \citealt{chiappini2008}, and references therein). 
 
\par We remark on the fact that \citet{romano2010} combine results of stellar models assuming only hydrostatic burning and rotation 
(Geneva group, for He, C, N, and O) with the results of models including explosive burning without rotation (Nomoto group, for heavier elements), giving rise to an 
inhomogeneous set of stellar yields, which can be physically incorrect. 
In the context of this study, the treatment of \citet{romano2010} has a marginal effect, 
since the metallicity is dominated by the oxygen and carbon contributions. 
On the other hand, \citet{nomoto2013} provide one of the most homogeneous set of stellar yields available at the present time, 
although it is still affected by the limitation of not including the effect of stellar rotation. 

\par In Figures \ref{fig:Oyields} and \ref{fig:Oyields_nomoto}, we show how the oxygen stellar yields of \citet{romano2010} and \citet{nomoto2013}, 
 respectively, vary as functions of the initial stellar mass and for different metallicities. 
The Geneva stellar yields are available only up $60\,\text{M}_{\sun}$ and we therefore assume in our standard case 
that the yields from $60$ to $100\,\text{M}_{\sun}$ are constant. On the other hand, the stellar yields of massive stars of 
\citet[included in \citealt{romano2010}, for the elements heavier than oxygen]{kobayashi2006} and \citet{nomoto2013} 
are available only up to $40\,\text{M}_{\sun}$ and thus we keep them constant for stars with larger initial mass. 
We remark on the fact that very massive stars are expected to leave a black hole as a remnant; 
therefore, a significant fraction of stellar nucleosythetic products in the ejecta of very massive stars may eventually fall back onto the black hole. 
This process might cause a reduction of the stellar yields of very massive stars. 

\par At solar metallicity, \citet{romano2010} include stellar yields which have been computed by applying a stellar rotational velocity $v_\mathrm{rot}=300$ km s$^{-1}$. 
From an observational point of view, \citet{ramirez2013} found that almost $80$ per cent of nearby stars rotate slower than  $v_\mathrm{rot}=300$ km s$^{-1}$, 
which thus can be considered as an approximate upper limit. To quantify the effect of stellar rotation in the stellar yields of oxygen from massive stars, 
in Table \ref{table_yields} we compare the predictions of models with and without stellar rotation, 
with the quantity $\langle P_{\mathrm{O}}\rangle$ being defined as 
the IMF-averaged yield of oxygen in the mass range $M=10$-$40\,\mathrm{M}_{\sun}$.  
The effect of stellar rotation in the Geneva stellar models is to increase 
the average oxygen stellar yield by a factor of $\sim1.5$ for stars with initial mass below 
$40\,\mathrm{M}_{\sun}$ (see also \citealt{hirschi2005}). Furthermore, at $\mathrm{Z}=4.0\times10^{-3}$,  the IMF-averaged oxygen stellar yield of 
\citet{nomoto2013} -- which neglect the effect of stellar rotation -- is larger than the value of the Geneva stellar models without stellar rotation, 
but rather similar to the corresponding value with rotation; 
conversely, at solar metallicity, the Geneva stellar models without rotation agree with \citet{nomoto2013}. 
IMFs containing a larger number of massive stars, such as the \citet{chabrier2003} and
\citet{kroupa2001} ones, amplify the oxygen enrichment of the ISM and give rise to larger values of $\langle P_{\mathrm{O}} \rangle$, whatever be the set of stellar 
yields assumed. 

\par In this article, we study the effect of different IMFs: the \citet{salpeter1955}, the \citet{kroupa1993}, the \citet{kroupa2001}, and the \citet{chabrier2003} IMFs, 
which are shown in Fig. \ref{fig:IMF} as normalized with respect to the \citet{salpeter1955} IMF. 
We have chosen these IMFs since they have been the most widely used by various authors. Moreover,
these IMFs give quite different weights to different stellar mass ranges, hence they will more clearly 
display differences in the final predicted net yields and return mass fractions. As one can notice from Fig. \ref{fig:IMF}, the \citet{kroupa1993} IMF predicts the largest fraction of intermediate mass stars, while having the lowest number of massive stars. On the other hand, the \citet{chabrier2003} and the \citet{kroupa2001} IMFs predict a higher number of both
intermediate-mass stars and massive stars than the \citet{salpeter1955} IMF.

\par Finally, we also explore the effect of the top-heavy IMFs of \citet{cappellari2012} and \citet{arimoto1987}, which are shown in Fig. \ref{fig:IMF_topheavy}, 
as normalized with respect to the \citet{salpeter1955} IMF. These IMFs are defined as a 
single-slope power law: $\phi(m)\propto m^{-(1+x)}$, with the \citet{cappellari2012} IMF having a slope $x=0.5$ and the \citet{arimoto1987} IMF assuming $x=0.95$. 
 
       %%%%%%%%%%%%%%%%%%%%%%%%%%%%%%%%%%%%%%%%%%
 
  \begin{figure}
\includegraphics[width=9.3cm]{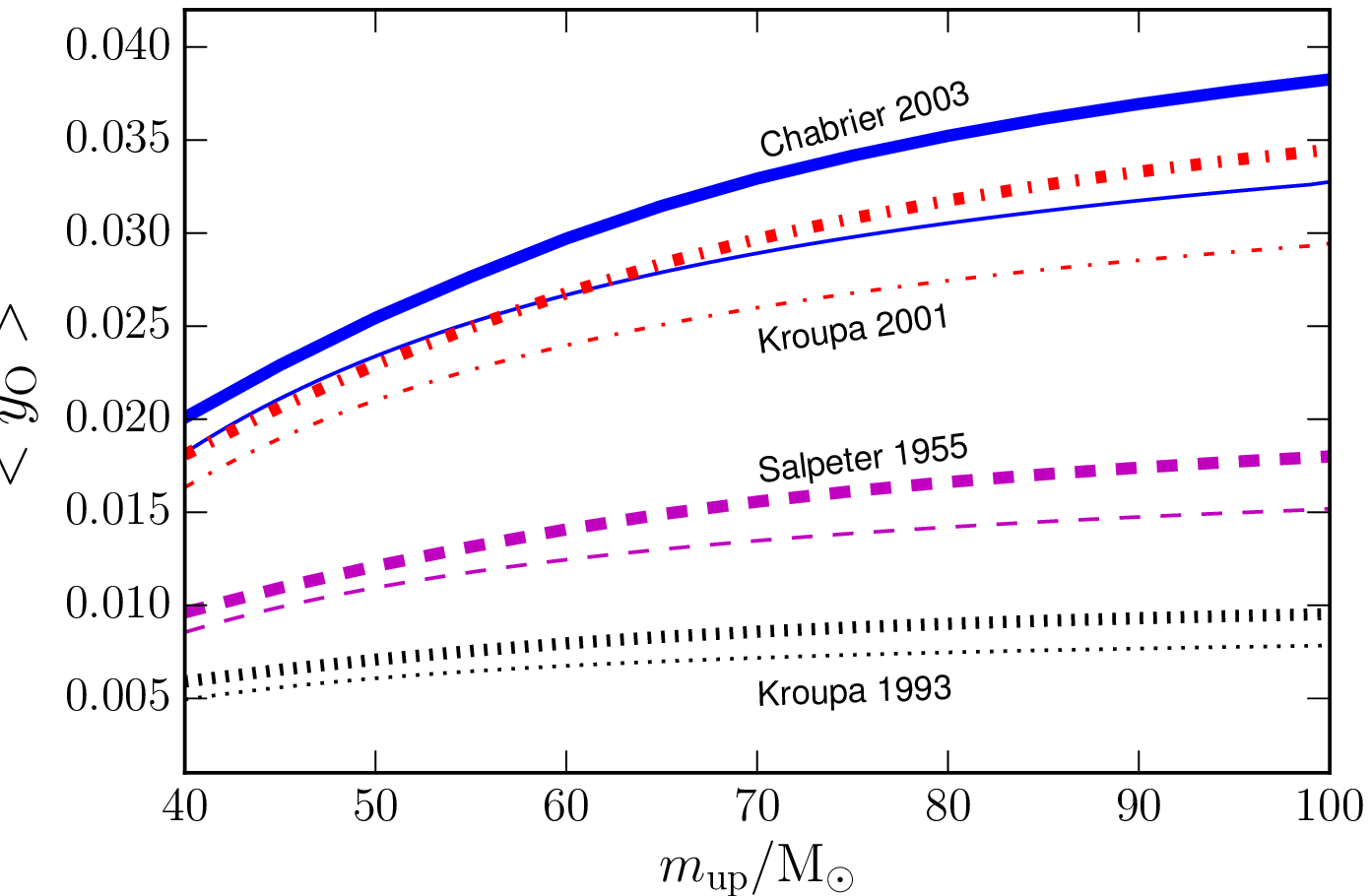}
\caption{In this figure, we show how $\langle y_{\text{O}} \rangle$ is predicted to vary as a function of $m_{\text{up}}$, which is defined 
as the upper mass cutoff 
of the IMF. 
We have computed $\langle y_{\text{O}} \rangle$, by averaging the net yield of oxygen over the metallicity 
range $1.0\times10^{-3}\le\text{Z}\le2.0\times10^{-2}$, within which $y_{\text{O}}$ turns out to be nearly constant. 
Thick lines represent our results for the \citet{romano2010} stellar yields, while thin lines represent the \citet{nomoto2013} stellar yields. 
The various curves with different colours correspond to the same IMFs as in Fig. \ref{fig:IMF}.  } 
     \label{fig:yO_trunc}
   \end{figure} 
   
%%%%%%%%%%%%%%%%%%%%%%%%%%%%%%%%%%%%%%%%%%

    %%%%%%%%%%%%%%%%%%%%%%%%%%%%%%%%%%%%%%%%%%

  \begin{figure}
\includegraphics[width=9.3cm]{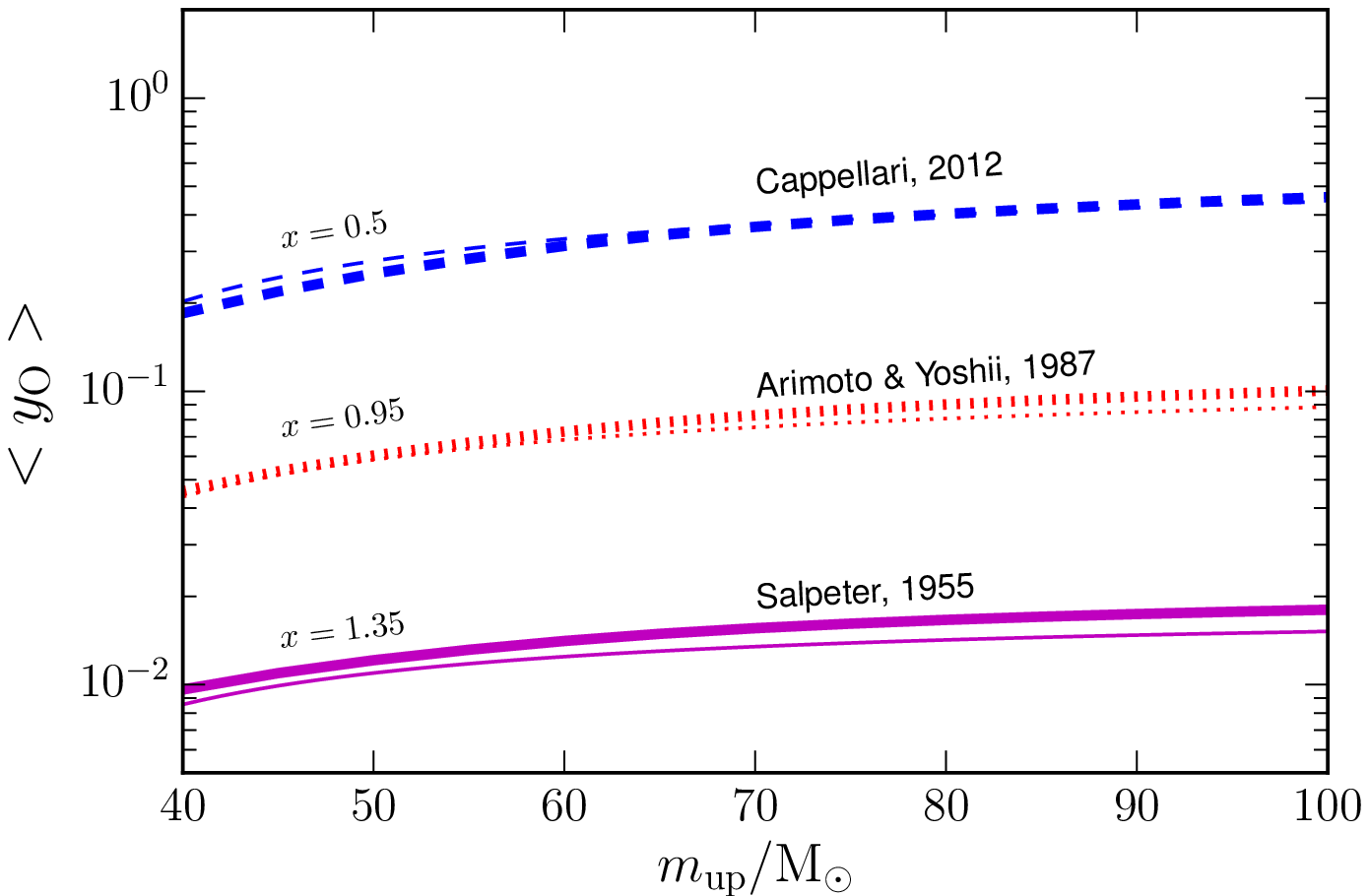}
\caption{In this figure, we report how the average net yield of oxygen, $\langle y_{\text{O}} \rangle$, varies as a function of the upper cutoff of the IMF, when assuming 
the top-heavy IMF of \citet[blue dashed line]{cappellari2012} and the IMFs of \citet[red dotted line]{arimoto1987} and \citet[solid line in magenta]{salpeter1955}. 
As in Fig. \ref{fig:yO_trunc}, $\langle y_{\text{O}} \rangle$ is computed by averaging $y_{\text{O}}$ over the metallicity range $1.0\times10^{-3}\le\text{Z}\le2.0\times10^{-2}$, 
and the various curves with different thickness correspond to the same stellar yields as in Fig. \ref{fig:yO_trunc}. }
     \label{fig:yO_topheavy}
   \end{figure} 
 
 %%%%%%%%%%%%%%%%%%%%%%%%%%%%%%%%%%%%%%%%%%

\section{Results} \label{results}
 
In this Section we present the net nucleosynthetic yields and return fractions obtained with the IMFs and
stellar yields discussed in the previous section. As mentioned above, we will focus on the yield of
oxygen, since it
is the element most commonly measured and taken as representative of the bulk of the metallicity, and
also
because it is an element for which the IRA approximation is appropriate. However,
we will provide a value also for the yield of the total mass of metals, although with some
cautionary warnings.

   %%%%%%%%%%%%%%%%%%%%%%%%%%%%%%%%%%%%%%%%%%

 \begin{table*}
 \caption[]{ \footnotesize{In this Table, we report the numerical values which we predict for the return mass fraction ($R$) and the yields of oxygen 
 and metals per stellar generation ($y_{\text{O}}$ and $y_{\text{Z}}$, respectively) as functions of the metallicity Z and for different IMFs. These values 
 have been computed by assuming that the upper mass cutoff of the IMF $m_\text{up}=100\,\text{M}_{\sun}$. The stellar yields are the ones of \citet{romano2010}.}}
 \begin{tabular}{c | c c c|c c c|c c c|c c c}
 \hline
  \multicolumn{13}{c}{Stellar yields: \citet{romano2010}} \\
   \small{Z} & \small{$R$} & \small{$y_{\text{O}}$}  & \small{$y_{\text{Z}}$} & \small{$R$} & \small{$y_{\text{O}}$} & \small{$y_{\text{Z}}$} & \small{$R$} & \small{$y_{\text{O}}$}  & \small{$y_{\text{Z}}$} & \small{$R$} 
 & \small{$y_{\text{O}}$} & \small{$y_{\text{Z}}$} \\
 \hline
 \rule{0pt}{3ex}
  & \multicolumn{3}{l}{IMF: \citet{salpeter1955}} & \multicolumn{3}{l}{IMF: \citet{chabrier2003}} & \multicolumn{3}{l}{IMF: \citet{kroupa1993}}  & \multicolumn{3}{l}{IMF: \citet{kroupa2001}} \\
\rule{0pt}{3ex}
 $1.0\times10^{-10}$ & $0.285$ & $0.028$ & $0.042$ & $0.436$ & $0.059$ & $0.088$ & $0.284$ & $0.015$ & $0.026$ & $0.411$ & $0.053$ & $0.079$  \\
\rule{0pt}{2ex}     
 $1.0\times10^{-5}$ & $0.285$ & $0.018$ & $0.028$ & $0.436$ & $0.039$ & $0.059$ & $0.284$ & $0.009$ & $0.017$ & $0.411$ & $0.035$ & $0.053$  \\
\rule{0pt}{2ex}    
 $5.0\times10^{-5}$ & $0.285$ & $0.018$ & $0.028$ & $0.436$ & $0.039$ & $0.059$ & $0.284$ & $0.009$ & $0.017$ & $0.411$ & $0.035$ & $0.053$  \\
\rule{0pt}{2ex}    
 $1.0\times10^{-4}$ & $0.285$ & $0.018$ & $0.029$ & $0.436$ & $0.039$ & $0.060$ & $0.284$ & $0.009$ & $0.018$ & $0.411$ & $0.035$ & $0.054$  \\
\rule{0pt}{2ex}    
 $5.0\times10^{-4}$ & $0.286$ & $0.018$ & $0.029$ & $0.437$ & $0.039$ & $0.060$ & $0.285$ & $0.009$ & $0.018$ & $0.412$ & $0.035$ & $0.054$  \\
\rule{0pt}{2ex}    
 $1.0\times10^{-3}$ & $0.286$ & $0.018$ & $0.029$ & $0.438$ & $0.039$ & $0.060$ & $0.287$ & $0.009$ & $0.017$ & $0.414$ & $0.035$ & $0.054$  \\
\rule{0pt}{2ex}    
 $5.0\times10^{-3}$ & $0.292$ & $0.018$ & $0.027$ & $0.447$ & $0.038$ & $0.057$ & $0.295$ & $0.009$ & $0.016$ & $0.422$ & $0.034$ & $0.051$  \\
\rule{0pt}{2ex}    
 $1.0\times10^{-2}$ & $0.295$ & $0.018$ & $0.028$ & $0.451$ & $0.038$ & $0.060$ & $0.299$ & $0.010$ & $0.017$ & $0.425$ & $0.034$ & $0.054$  \\
\rule{0pt}{2ex}    
 $2.0\times10^{-2}$ & $0.298$ & $0.018$ & $0.031$ & $0.455$ & $0.037$ & $0.065$ & $0.302$ & $0.010$ & $0.018$ & $0.430$ & $0.034$ & $0.059$  \\ 
 \hline
\end{tabular}
\label{table1}
\end{table*} 

 %%%%%%%%%%%%%%%%%%%%%%%%%%%%%%%%%%%%%%%%%%  
   
\par In Table \ref{table1}, we show how the net yield of oxygen per stellar generation varies as a function of
the IMF and metallicity.  In our ``fiducial'' case, reported in Table \ref{table1}, we assume $m_{\text{up}}=100\,\text{M}_{\sun}$. 
Concerning the dependence on metallicity, the most interesting result is that
the yield $y_{\text{O}}$ is roughly constant down to very low metallicities.
This result implies that the assumption of a time-independent net oxygen yield, as generally treated in analytical models, is a reasonable one. 
Interestingly, 
we find an enhancement of $y_{\text{O}}$ for metal-free stellar populations (case with $Z=1.0\times10^{-10}$ from \citealt{ekstrom2008}). 
In fact, it is well established that, at very low Z, the mixing induced by rotation is particularly efficient \citep{chiappini2008}; in this way, 
the nucleosynthetic products of the $3\alpha$ reaction in the inner He-burning zone can diffuse to the outer stellar zones, so that  
radiative winds and mass loss (boosted by the high surface enrichment in heavy elements) are highly enriched with the CNO elements; 
this cannot be obtained by models of metal-free non-rotating massive stars (see, for example, \citealt{maeder2009}, for a detailed discussion).

\par On the other hand, $y_{\text{O}}$ is strongly dependent on the assumed IMF. The highest oxygen
yield is obtained when adopting a \citet{chabrier2003} IMF, because this particular IMF contains the
largest number of massive stars compared to the other IMFs explored in this paper (see Fig. \ref{fig:IMF}). 
The IMF of \citet{kroupa1993}, instead, predicts the lowest $y_{\text{O}}$, since it contains the lowest fraction of high mass stars. 
In this context, it is important to distinguish the two IMFs suggested by Kroupa. 
In fact the \citet{kroupa2001} is very similar to the IMF of \citet{chabrier2003} and predicts a substantially higher yield than \citet{kroupa1993}. 
The \citet{salpeter1955} IMF predicts a net yield roughly halfway between the \citet{chabrier2003} and \citet{kroupa1993} IMFs.

\par In Table \ref{table1}, we show also how $y_{\text{Z}}$ (where Z here is the sum of all metals) is predicted to vary as a function of the 
different IMFs and metallicities. 
These values are shown here only for reference with previous works attempting to model the total metal content of galaxies, 
however we caution the reader against a blind application of analytical models assuming the IRA to the total metal content. 
Finally, in the same Table, we report the values of the returned fraction $R$, which is rather constant as a function of metallicity but shows some change for different IMFs. 
The approximate constancy of $R$ as a function of the metallicity is due to the fact that this quantity is strongly dominated by the 
H and He contributions. 

\par Our results for $y_{\text{O}}(\text{Z}) $, $y_{\text{Z}}(\text{Z}) $ and $R(\text{Z})$, as obtained with the \citet{nomoto2013} set of stellar yields, are reported 
in Table \ref{table_nomoto}.  The effect of the various IMFs is the same as discussed above for the stellar yields of \citet[see Table \ref{table1}]{romano2010}. 
On the other hand, by comparing the predicted net yields of metals and oxygen of \citet{romano2010} with the ones of \citet{nomoto2013}, 
we can quantify the uncertainty introduced by different input stellar yields by a factor which is $\sim1.5$.

 %%%%%%%%%%%%%%%%%%%%%%%%%%%%%%%%%%%%%%%%%%

 \begin{table*}
 \caption[]{ \footnotesize{In this Table, we report the numerical values which we predict for the return mass fraction ($R$) and the yields of oxygen 
 and metals per stellar generation ($y_{\text{O}}$ and $y_{\text{Z}}$, respectively) as functions of the metallicity Z and for different IMFs. These values 
 have been computed by assuming that the upper mass cutoff of the IMF $m_\text{up}=100\,\text{M}_{\sun}$. The stellar yields are the ones of \citet{nomoto2013}.}}
 \begin{tabular}{c | c c c|c c c|c c c|c c c}
 \hline
   \multicolumn{13}{c}{Stellar yields: \citet{nomoto2013}} \\
 \small{Z} & \small{$R$} & \small{$y_{\text{O}}$} & \small{$y_{\text{Z}}$} & \small{$R$} & \small{$y_{\text{O}}$} & \small{$y_{\text{Z}}$} &  \small{$R$} & \small{$y_{\text{O}}$} 
 & \small{$y_{\text{Z}}$} & \small{$R$} & \small{$y_{\text{O}}$} & \small{$y_{\text{Z}}$}  \\
 \hline
 \rule{0pt}{3ex}
  & \multicolumn{3}{l}{IMF: \citet{salpeter1955}} & \multicolumn{3}{l}{IMF: \citet{chabrier2003}} & \multicolumn{3}{l}{IMF: \citet{kroupa1993}}  & \multicolumn{3}{l}{IMF: \citet{kroupa2001}} \\
\rule{0pt}{3ex}
 $0.0$ & $0.261$ & $0.021$ & $0.043$ & $0.403$ & $0.044$ &  $0.087$ & $0.244$ & $0.011$ &  $0.024$ & $0.380$ & $0.040$  & $0.079$  \\ 
 \rule{0pt}{2ex}
 $1.0\times10^{-3}$ & $0.293$ & $0.018$ &  $0.026$ & $0.450$ & $0.038$ &  $0.055$ & $0.291$ & $0.009$ &  $0.014$ & $0.424$ & $0.034$  & $0.050$  \\
\rule{0pt}{2ex}     
 $5.0\times10^{-3}$ & $0.300$ & $0.016$ &  $0.025$ & $0.459$ & $0.034$ &  $0.052$ & $0.300$ & $0.008$ &  $0.013$ & $0.433$ & $0.030$ &  $0.047$ \\
\rule{0pt}{2ex}    
 $1.0\times10^{-2}$ & $0.302$ & $0.015$ &  $0.024$ & $0.463$ & $0.032$ &  $0.051$ & $0.303$ & $0.008$ &  $0.013$ & $0.436$ & $0.029$ &  $0.046$ \\
\rule{0pt}{2ex}    
 $2.0\times10^{-2}$ & $0.305$ & $0.014$ &  $0.023$ & $0.466$ & $0.030$ &  $0.049$ & $0.307$ & $0.007$ &  $0.012$ & $0.439$ & $0.027$ &  $0.044$  \\
\rule{0pt}{2ex}  
 $5.0\times10^{-2}$ &$0.304$ & $0.017$ &  $0.023$ & $0.466$ & $0.036$ &  $0.049$ & $0.307$ & $0.009$ &  $0.012$ & $0.439$ & $0.032$ &  $0.044$  \\    
 \hline
\end{tabular}
\label{table_nomoto}
\end{table*} 

 %%%%%%%%%%%%%%%%%%%%%%%%%%%%%%%%%%%%%%%%%%   

\par In Fig. \ref{fig:yO_trunc}, we  explore how the choice of the upper cutoff of the IMF, $m_{\text{up}}$, affects the average net yield of oxygen, where 
$\langle y_{\text{O}} \rangle$ stands for the net yield of oxygen as averaged in the metallicity range $1.0\times10^{-3}\le\text{Z}\le2.0\times10^{-2}$. 
Our results are shown for different IMFs (different colors) and different stellar yield compilations (thick and thin lines represent 
our results with \citealt{romano2010} and \citealt{nomoto2013}, respectively). 
By looking at the figure, we see that the difference between the
case with $m_{\text{up}}=100\,\text{M}_{\sun}$ and $m_{\text{up}}=40\,\text{M}_{\sun}$ is as much as a factor of about
two. The difference  is only $\sim50$ per cent between $m_{\text{up}}=100\,\text{M}_{\sun}$ and $m_{\text{up}}=60\,\text{M}_{\sun}$. 
We find that, when assuming the \citet{chabrier2003} and \citet{kroupa1993} IMFs, 
the differences in $\langle y_{\text{O}} \rangle$ with different upper mass limits are almost doubled and halved, respectively, 
with respect to the case with the \citet{salpeter1955}. 
 Fig. \ref{fig:yO_trunc} shows that the global variation of $\langle y_{\text{O}} \rangle$ spanned by all ``classical'' IMFs and the possible range of 
 $m_{\text{up}}$ is nearly a factor of ten. 
 \par By looking at Fig. \ref{fig:yO_trunc}, the curves corresponding to the \citet{romano2010} stellar yields lie always above the curves with the  
 \citet{nomoto2013} stellar yields. This difference enlarges as $m_{\text{up}}$ increases, because \citet{romano2010} provide the oxygen stellar yields up to 
 $60\,\text{M}_{\sun}$, while \citet{nomoto2013} only up to $40\,\text{M}_{\sun}$ (see also Fig. \ref{fig:Oyields}), and we keep these stellar yields constant for 
 stars with larger initial stellar mass. 
 The latter assumption can introduce a systematic effect in the final values 
of $y_\text{O}$. We find that, by varying the upper limit of the integral at the numerator of equation \ref{eq:y} and by normalizing the IMF up to 
$m_\text{up}=100\,\text{M}_{\sun}$, the trend of the resulting $\langle y_{\text{O}} \rangle$ is similar to the trend of $\langle y_{\text{O}} \rangle$ as a function of the 
upper cutoff of the IMF (see Fig. \ref{fig:yO_trunc}). 
 
\par Assuming a top-heavy IMF can cause an even larger increase of the yield of oxygen per stellar generation, being larger the number of massive stars which are 
present. We explore the effect of two top-heavy IMFs \citep{cappellari2012,arimoto1987}, both defined as a single-slope power law. 
Our results for the $\langle y_{\text{O}} \rangle$ vs. $m_\mathrm{up}$ relations are shown in Fig. \ref{fig:yO_topheavy} for different IMFs and stellar yield assumptions. 
By looking at Fig. \ref{fig:yO_topheavy}, as the slope of the IMF is decreased from $x=1.35$  \citep{salpeter1955} down to $x=0.95$ \citep{arimoto1987}  
and $x=0.5$ \citep{cappellari2012}, the IMF becomes top-heavier and the value of $y_\mathrm{O}$ becomes larger and larger; furthermore, 
the standard deviation of $\langle y_{\text{O}} \rangle$ becomes larger as the slope $x$ is decreased, indicating that $\langle y_{\text{O}} \rangle$ is slightly 
more influenced by the metallicity-dependence of the assumed set of stellar yields. 

\par The predicted values of $y_\mathrm{O}$ for single-slope top-heavy 
IMFs are very high and they may either imply that a top-heavy IMF star formation mode 
has only lasted for a short interval of the galaxy lifetime and not relevant 
for the integrated metal production, or that a single power-law is not a proper representation of the ``top-heavy IMF'' and that a broken power-law is a more proper 
description. 
   
\par There is increasing evidence in the literature that the IMF may vary among different types of stellar systems, such as spheroids and disk galaxies, as well as faint dwarf galaxies 
(e.g. see \citealt{cappellari2012,conroy2012,weidner2013}). 
Observationally, a \citet{kroupa1993} IMF is favoured in describing the chemical evolution of the solar vicinity (see \citealt{romano2010}), 
and the disk of spirals similar to the Milky Way, while the \citet{kroupa2001} and \citet{chabrier2003} IMFs are probably better for describing the evolution of spheroids such as bulges and ellipticals 
(see, for example, \citealt{chabrier2014}). We have shown that the range of commonly adopted IMFs 
(even neglecting the extreme top-heavy IMFs) implies a large variation of net yield per stellar generation. 
This fact could add an extra systematic to studies attempting to model the observed abundances, which should be taken into
account by properly using our compilation of yield for the different classes of galaxies.  
More specifically, if the IMF is not universal, we can expect a difference in net yield up to a factor larger than three for classical, widely-used IMFs, 
and even much larger for top-heavy IMFs.

\section{Conclusions and discussion} \label{conclusions}

The blooming of extensive spectroscopic surveys of local and distant galaxies have fostered the
use of metallicities to constrain the star formation history, feedback processes (outflows) and
gas inflows across the cosmic epoch, by comparing the observations with the expectations of analytical
and numerical models. One key element in such a comparison is the yield per stellar generation,
which is often assumed as a fixed value. 
In this article, we have used a numerical code of chemical evolution, which includes modern stellar nucleosynthetic yields, 
to explore the effect of the metallicity and IMF on the net yield of oxygen per stellar generation 
and on the return mass fraction; our results have shown that the yield can change by 
large factors. Therefore, if the yield associated with the appropriate IMF is not used, this can
produce inconsistent results and large systematic errors. 

\par We have provided results for two different sets of stellar 
yields, which are \citet{romano2010} and \citet{nomoto2013}. The former is an inhomogeneous set, since it combines results 
of hydrostatic burning in rotating massive stars  (He, C, N, and O from Geneva stellar models) with results of explosive burning without stellar rotation 
(heavier elements from \citealt{kobayashi2006}), 
which is -- in principle -- an incorrect treatment. 
On the other hand, \citet{nomoto2013} is a homogeneous set of stellar yields, with the only limitation of 
not including the effect of stellar rotation, expected to have a strong impact at the very low metallicities. 
We have found that the uncertainty introduced by assuming different sets of stellar yields can be quantified by $\sim0.2\,\text{dex}$. 

\par The effect of assuming different IMFs can cause large differences in the net oxygen yield. We have found that the \citet{kroupa2001} and \citet{chabrier2003} predict the highest oxygen yield, 
roughly a factor of two higher than for a \citet{salpeter1955} IMF.
On the other hand, by assuming a \citet{kroupa1993} IMF, we obtain the smallest net yield, roughly a factor of two lower than for a \citet{salpeter1955} IMF.

\par The yield per stellar generation also depends significantly on the upper mass cutoff of the IMF. The differences 
between the case with $m_{\text{up}}=100\,\text{M}_{\sun}$ and $m_{\text{up}}=40\,\text{M}_{\sun}$ are of the order of
a factor of two with the \citet{salpeter1955}. Since the IMF of \citet{chabrier2003} predicts a larger number 
of massive stars, that difference is doubled with this IMF, whereas it is halved with the \citet{kroupa1993} IMF, which predicts the lowest number of massive stars. 
If one takes into account both the variation with IMF shape and upper stellar mass cutoff, {\it the variation of the
yield per stellar generation can span more than a factor of ten.}

\par We note that populations of highly enriched galaxies -- whose 
metallicities were deemed uncomfortably high -- can be easily explained by means of a large yield per stellar generation, as the one 
associated with commonly used IMFs, such as \citet{chabrier2003} and \citet{kroupa2001}. 
Similarly, our results should warn about a proper use of the so-called {\it effective yield},
$y_{\text{eff}}=\text{Z}/\ln \left( \mu^{-1} \right)$, which is observationally derived by inverting equation (\ref{eq:simple}). 
In particular, the finding of $y_{\text{eff}}<y_{\text{Z}}$
is generally modelled in terms of enriched outflows, inflow of pristine gas, or both (e.g.
\citealt{tremonti2004,erb2008,mannucci2009,troncoso2014}), whereas the finding
of $y_{\text{eff}}>y_{\text{Z}}$ is sometimes used as an indicator of inaccurate metallicity measurements or
inappropriate metallicity calibrations. We conclude that the deviation of $y_{\text{eff}}$ from a ``fiducial'', ``true'' yield
also may be partly associated with IMF being different than assumed. For the same reason, high values of the the
effective yield (and in particular $y_{\text{eff}} > y_{\text{Z}}$)  may be indicative of an IMF favouring massive stars \citep{chabrier2003,kroupa2001} 
and/or of an high mass cutoff of the IMF itself. 
 By assuming single-slope top-heavy IMFs, such as 
the ones proposed by \citet{arimoto1987} or \citet{cappellari2012}, we find very high values for the yields of oxygen per stellar generation, 
which are also slightly influenced by the metallicity-dependence of the stellar yields.  
 
\par The dependence on metallicity of the yield is reassuringly small. A significant variation is only found at
extremely low metallicities ($\text{Z}=1.0\times10^{-10}$). Although the latter result may be an important aspect to take into account for models
of primordial galaxies, it strongly relies on the assumed set of stellar yields, which are particularly affected by 
uncertainties at extremely low metallicities.

\par We remind the reader that IRA provides correct results only for chemical elements restored 
into the ISM on short typical time-scales; oxygen represents the best example of such a chemical element, since it is also the most 
abundant metal by mass in the Universe. On the other hand, analytical models working under IRA assumption 
fail in following the evolution of chemical elements produced by long-lifetime sources; examples of such chemical elements are 
carbon, nitrogen and iron. Hence we warn the reader against a blind application of IRA 
for the total mass of metals. In order to take into account the lifetimes of the various stellar producers in detail, 
numerical models of chemical evolution should be used.  

\par Our compilation of numerical values of the yield per stellar generation, for different IMFs, different upper mass cutoffs and
different metallicities, will hopefully be useful to properly investigate the metallicity in galaxies across cosmic epochs, 
by tackling one of the (generally not acknowledged) major sources of uncertainty.

\section*{Acknowledgements}

FM acknowledges financial support from PRIN-MIUR~2010-2011 project 
`The Chemical and Dynamical Evolution of the Milky Way and Local Group Galaxies', prot.~2010LY5N2T. 
We thank an anonymous referee for his/her constructive comments.

\bsp

\label{lastpage}

\end{document}